\documentclass[]{spie}  

\usepackage{amsmath,amsfonts,amssymb}
\usepackage{siunitx}
\usepackage{graphicx}
\usepackage{subcaption}
\usepackage{textcomp}
\usepackage[colorlinks=true, allcolors=blue]{hyperref}

\title{Viscoelastic Characterization of Melanoma Cells Using Brillouin Spectroscopy}

\author[a]{Mykyta Kizilov}
\author[a]{Vsevolod Cheburkanov}
\author[a]{Sujeong Jung}
\author[a]{Vladislav V. Yakovlev}

\affil[a]{Department of Biomedical Engineering, Texas A\&M University, College Station, TX, 77843, USA}

\authorinfo{Further author information: (Send correspondence to MK)\\E-mail: mykyta.kizilov@tamu.edu, Telephone: 1 979 229 4349\\}

\begin{document} 
\maketitle

\begin{abstract}
In this study, Brillouin spectroscopy was employed to investigate the viscoelastic properties of melanoma cells in vitro. Using a custom-built confocal Brillouin microspectrometer, we obtained Brillouin shifts and full width at half maximum (FWHM) values, enabling the non-invasive assessment of cellular stiffness and viscosity. The Brillouin spectra revealed the biomechanical characteristics of melanoma cells, with measured shifts and FWHM values providing a detailed viscoelastic profile. These findings demonstrate the capability of Brillouin microscopy to probe the mechanical properties of cancer cells at the subcellular level. This technique holds significant potential for advancing cancer research by providing insights into the mechanical behavior of melanoma cells, which could inform the development of diagnostic tools and therapeutic strategies based on cellular biomechanics.
\end{abstract}


\section{Introduction and Objective}
Historically, Brillouin spectroscopy has been employed to assess the mechanical properties of various cell lines and tissues \cite{2025CheburkanovBrillouinSPIE}. The system utilized in the present study is optimized for the efficient acquisition of both low-resolution stiffness maps of highly scattering samples, such as cell-laden hydrogels and scar tissue, as well as high axial resolution elasticity maps of individual cells \cite{2025CheburkanovGliaSPIE, troyanova2019differentiating}.

The primary objective of this study was to evaluate the feasibility of Brillouin spectroscopy for imaging melanoma cells. During the experimental process, the authors observed a significant phenomenon of cell death at elevated laser power levels. However, by adjusting the exposure parameters and, consequently, the dwell time of the laser, the authors successfully enhanced cell viability under high-intensity illumination.

This observation provides a foundation for the secondary objective of the study, which aims to investigate the mechanical effects of laser-induced damage on the cells \cite{Wetzel2011Single}.
\label{sec:intro}

\section{Experiment design}
Our collaborators supplied melanoma cell lines cultured in tissue flasks for imaging purposes \cite{meeth2016yumm}. The flasks were securely positioned on the microscope stage and remained fixed throughout all experimental manipulations conducted during the day. Upon completion of the imaging sessions, the cells were returned to a 37$^{\circ}$C incubator to facilitate recovery.

The microscope objective employed in this study was the Nikon Fluor CFI60 60X water-dipping objective (numerical aperture, NA 1.0). This objective was infinity-corrected for 0.17 mm thick cover glass; therefore, variations in flask material thickness had a negligible impact on the effective NA for our applications.

All samples were imaged within cell culture flasks, where the cells were adhered to the bottom and submerged in nutrient media. The initial imaging run was designed as a test study to evaluate the capability of the system to accurately capture images of cells in this configuration, comparing the quality of widefield and confocal Brillouin channels' focal plane matching. Approximately 10mW of 532nm laser power was delivered to the sample.

Spatial sampling was varied between 1 $\mu$m and 2 $\mu$m to investigate the phototoxic effects of high-power 532 nm radiation on the cells \cite{gottschalk2015short}. The field of view was scanned in a consistent 60 by 60 grid pattern for all samples. The exposure time for each measurement was maintained between 30 and 40 ms.

\subsection{Setup description and acquisition settings}

\begin{figure*}[h]
    \centering
    \includegraphics[width=1\linewidth]{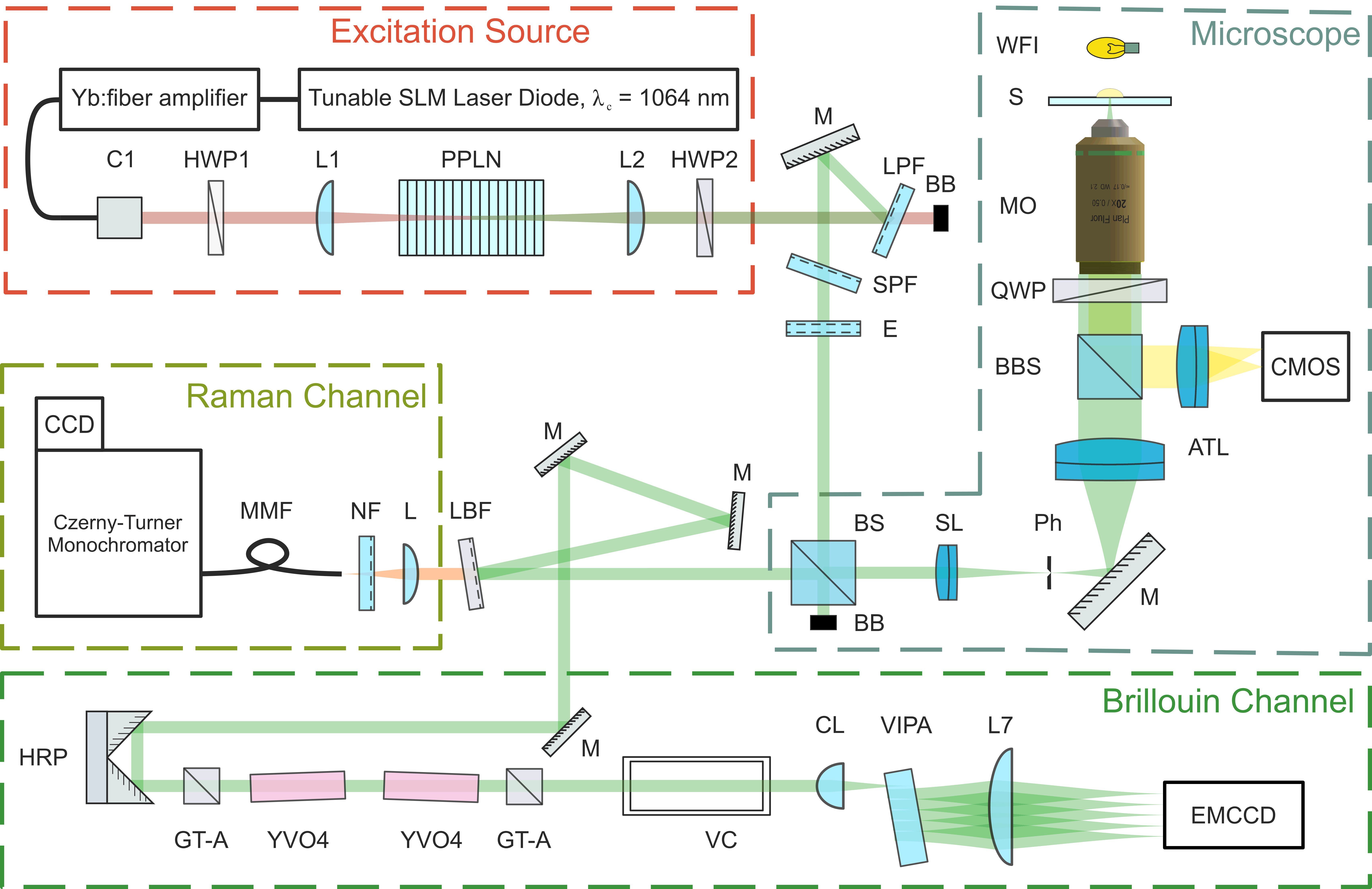}
    \caption{Brillouin confocal microspectrometer layout. \textbf{C}: fiber collimator, \textbf{HWP}: half-wave plate, \textbf{PPLN}: periodically poled lithium niobate second harmonic generation crystal, \textbf{BB}: beam block, \textbf{LPF}: long-pass filter, \textbf{M}: mirror, \textbf{SPF}: short-pass filter, \textbf{E}: Fabry-Perot etalon, \textbf{BS}: polarizing beamsplitter cube, \textbf{SL}: achromatic doublet scan lens, \textbf{Ph}: precision pinhole, \textbf{ATL}: achromatic tube lens (200mm EFFL), \textbf{BBS}: broadband beamsplitter in quick-insert mount, \textbf{QWP}: quarter-wave plates, \textbf{MO}: microscope objective lens, \textbf{S}: sample, \textbf{WFI}: widefield attachment illumination, \textbf{CMOS}: 5MP camera with a complimentary metal-oxide-semiconductor sensor, \textbf{LBF}: 532nm Bragg grating notch filter, \textbf{NF}: hard-coated interference notch filter, \textbf{MMF}: single core multimode fiber, \textbf{CCD}: camera with a charge-coupled device sensor, \textbf{HRP}: hollow-roof prism, \textbf{GT-A}: Glan-Taylor polarizer, \textbf{YVO4}: non-doped yttrium orthovanadate crystal, \textbf{VC}: Iodine vapor cell, \textbf{CL}: cylindrical lens, \textbf{VIPA}: Virtually Imaged Phase Array, \textbf{EMCCD}: CCD with electron multiplying amplification capabilities.}
    \label{fig:1}
\end{figure*}

The schematic of this setup is depicted in Figure \ref{fig:1}. The system comprises four primary subassemblies: an excitation source, a microscope, a confocal pinhole assembly, and the custom-built Brillouin microspectrometer.

\subsubsection{Excitation}

A custom-built 532 nm laser with a linewidth of less than 1 MHz was utilized as the excitation source. This wavelength was produced as the second harmonic of 1064 nm laser radiation within a periodically poled LiNbO3:MgO crystal (Covesion Ltd.). The 1064 nm light was generated by a tunable single longitudinal mode laser diode (Koheras Adjustik Y10, NKT Photonics) and subsequently amplified using an Yb-doped fiber amplifier (Koheras Boostik HPA Y10, NKT Photonics).

The desired wavelength of 532 nm was effectively isolated from the residual 1064 nm pump light through the use of a combination of long-pass (LPF) and short-pass (SPF) filters. Additionally, a Fabry-Perot etalon, characterized by a free spectral range of 30 GHz and a finesse exceeding 30, was incorporated into the excitation pathway to serve as a sub-GHz laser cleanup filter. This configuration enhanced the purity of the excitation light by minimizing unwanted spectral components, thereby optimizing the overall performance of the system. 

\subsubsection{Microscope}

The microscope body was designed and constructed using standard optomechanical components (Thorlabs). The laser power delivered to the sample was regulated through the use of a 532 nm half-wave plate in conjunction with a polarizing beamsplitter cube (BS, Thorlabs PBS251). A polymer quarter-wave plate (QWP) was employed to achieve orthogonal polarization between the incident and scattered light, transmitting the scattered light through the BS towards the spectrometer for Brillouin and Raman \cite{2025HarringtonDUVChemPhysChem, 2025HarringtonDUVSPIE, 2025KizilovRamanSPIE} channel detection.

To ensure optimal beam expansion, the laser beam was enlarged to a diameter of 10 mm at $1/e^2$ using a Keplerian telescope beam expander. A precision pinhole was positioned at the intermediate image plane of the telescope, serving to clean up the excitation beam and provide spatial filtration of the collected signal. The microscope was designed to be compatible with the Nikon CFI60 objectives, utilizing a tube lens (ATL) with a focal length of 200 mm. The microscope objective (MO) delivered approximately 10 ± 0.3 mW of 532 nm radiation to the sample while simultaneously collecting the scattered photons.

Target acquisition within the field of view was facilitated by a camera (CMOS) coupled to the microscope with a 200 mm achromatic lens. Transparent specimens were illuminated in transmission mode using a fiber-coupled 6000K LED (Mightex), designated as WFI in Fig. \ref{fig:1}. The position of the sample within the field of view was precisely adjusted via a microscope stage equipped with sub-micron precision. A MCL Nano-LPS stage was used for fine and a custom MCL MicroStage for coarse positioning (Mad City Labs).

\subsubsection{Confocal pinhole selection}

The confocal pinhole was selected to transmit no more than 1 Airy Unit (AU), thereby leveraging the advantages of confocal microscopy, including optical sectioning of the sample and isolation of data from individual layers. To validate pinhole size selection, a broadband dielectric mirror was precisely positioned in the object plane of the MO, and its axial position was systematically scanned. During this process, we measured the throughput of the pinhole and visually assessed beam quality.

Our measurements indicated that at the peak approximately 75\% of the signal was transmitted through the chosen pinhole, and the observed light pattern closely resembled the Fraunhofer diffraction pattern on the power meter. This observation suggests that the selected pinhole diameter was in fact less than 1 AU for the given system configuration, confirming it as a suitable design choice.

\subsubsection{Custom-built Brillouin spectrometer}

The experimental setup involved the detection of both Raman and Brillouin scattered photons, which were effectively separated using a Bragg volumetric grating notch filter optimized for a 532 nm excitation wavelength. This approach facilitated the distinct separation of Brillouin and Raman signals, while also enabling verification of system alignment and beam collimation, with the latter being influenced by the underlying principles of the Bragg grating design and operation.

The filtered signal was coupled into the entrance pupil of a custom-designed Brillouin spectrometer. To mitigate the influence of elastically scattered photons, an iodine vapor cell (Thorlabs) was employed, maintained at a temperature of  $70 ^\circ C$. By implementing a double-pass beam propagation configuration and precisely tuning the laser output wavelength to coincide with the strongest absorption band of molecular iodine, suppression ratios exceeding 40 dB were achieved. The optimal absorption wavelength was determined to be 531.9363 nm, corresponding to line 638, with a wavenumber of 18,799.244 $cm^{-1}$. 

In instances where iodine suppression was inadequate, a narrow-band polarization filter was incorporated into the system. This filter configuration utilized two Glan-Taylor (GT-A) prisms serving as a polarizer and analyzer, respectively, along with a pair of yttrium orthovanadate crystals functioning as long polarization rotators. The dimensions and alignment of the crystals were meticulously adjusted to minimize the transmission of elastically scattered photons by rotating the polarization plane of the elastically scattered signal to a configuration orthogonal to that of the analyzer.

The Brillouin scattering signal was analyzed using a high-dispersion, custom-built single-stage VIPA spectrometer. The VIPA (OP-6721-3371-2, Lightmachinery Inc.) was specifically optimized for 532 nm and featured a free spectral range of 29.98 GHz (1 $cm^{-1}$). Signal spectra were recorded with a water-cooled EMCCD camera (Andor Newton 970P, Oxford Instruments).

\section{Data analysis}
\subsection{Signal Pre-Processing}
\begin{figure}[!h]
    \centering
    \includegraphics[width=0.6\linewidth]{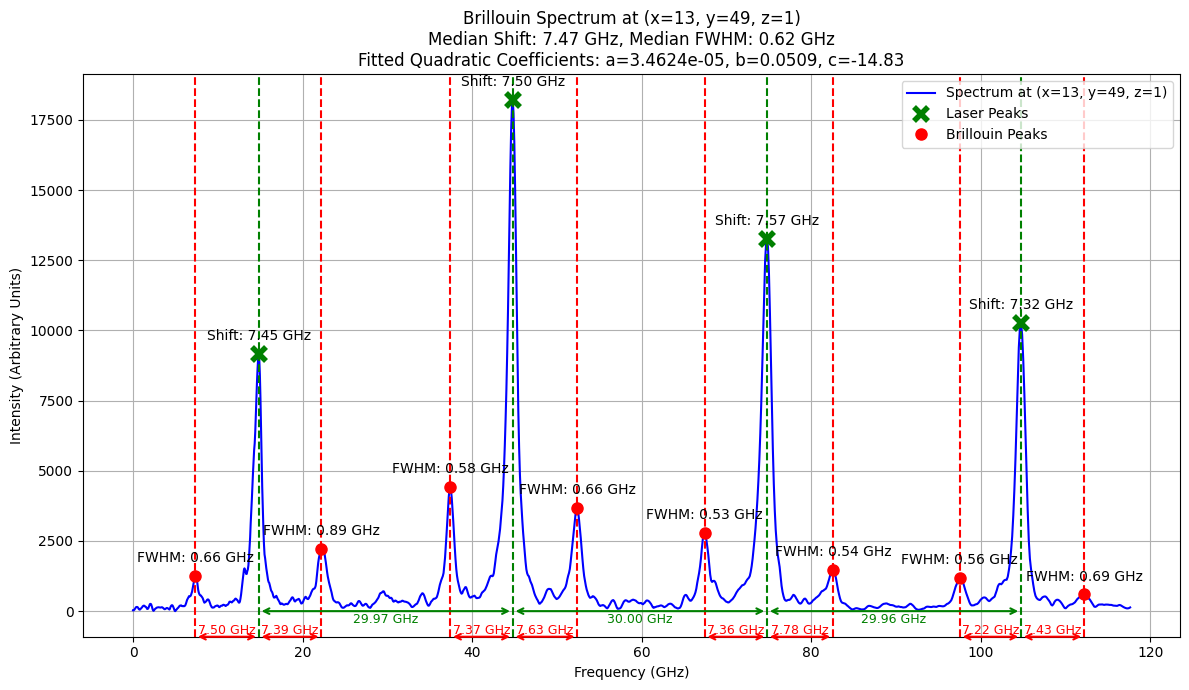}
    \caption{Brillouin spectrum}
    \label{fig:Brillouin_spectrum}
\end{figure}

For each pixel in each layer, the Brillouin spectrum was recorded. 
After recording each spectrum, we followed the next steps:
\begin{enumerate}
    \item The raw signal from the spectrometer's CCD was processed using a Savitzky-Golay filter \cite{savitzky1964smoothing} to enhance signal quality. The optimal signal-to-noise ratio (SNR) of the spectra was obtained with a filter window width of 7 and a fitted polynomial order of 3.
    \item The positions of specific spectral lines resulting from Rayleigh (elastically) scattered photons, as well as Stokes and Anti-Stokes Brillouin shifted lines, were precisely determined.

    \item Due to the non-linear spectral response of the Virtually Imaged Phase Array (VIPA), it was necessary to compute the dispersion curve of the spectrometer to accurately convert the observed shift values from the linear domain on the CCD sensor into the frequency domain. The dispersion curve was derived from the spectra corresponding to each measurement point, utilizing the fixed positions of spectral lines from Rayleigh (elastically) scattered photons. Given the operational principles of VIPA, we can determine the exact separation between these peaks in the frequency domain, which is established at 29.98 GHz.
    
    \item The values of Brillouin shift and Brillouin peaks Full Width Half Maximum (FWHM) are extracted from the spectra.
\end{enumerate}
The processed signal for a single pixel is illustrated in Fig. \ref{fig:Brillouin_spectrum}

\begin{figure}[!h]
  \centering
  \includegraphics[width=0.2\textwidth]{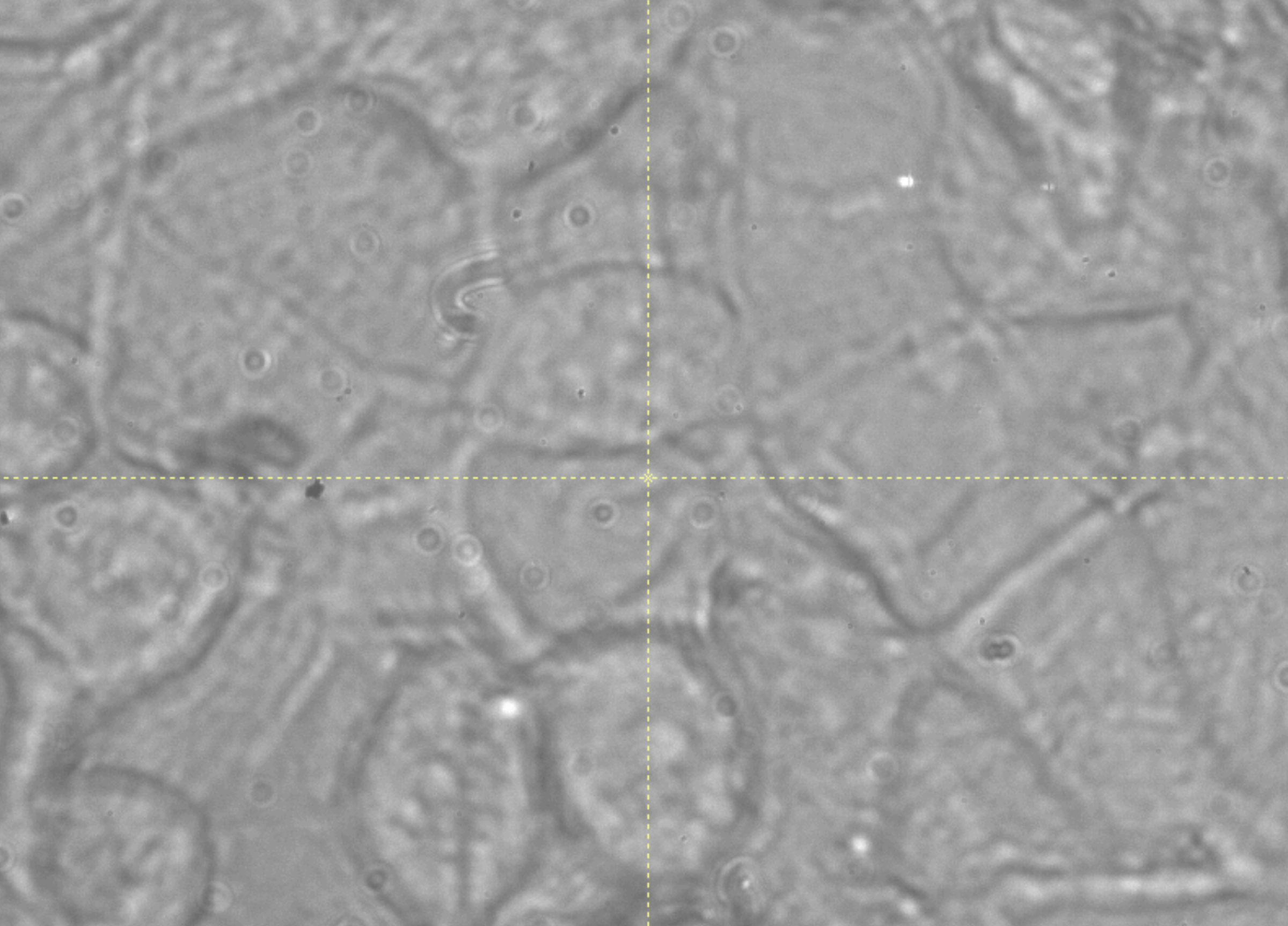}
  \includegraphics[width=0.2\textwidth]{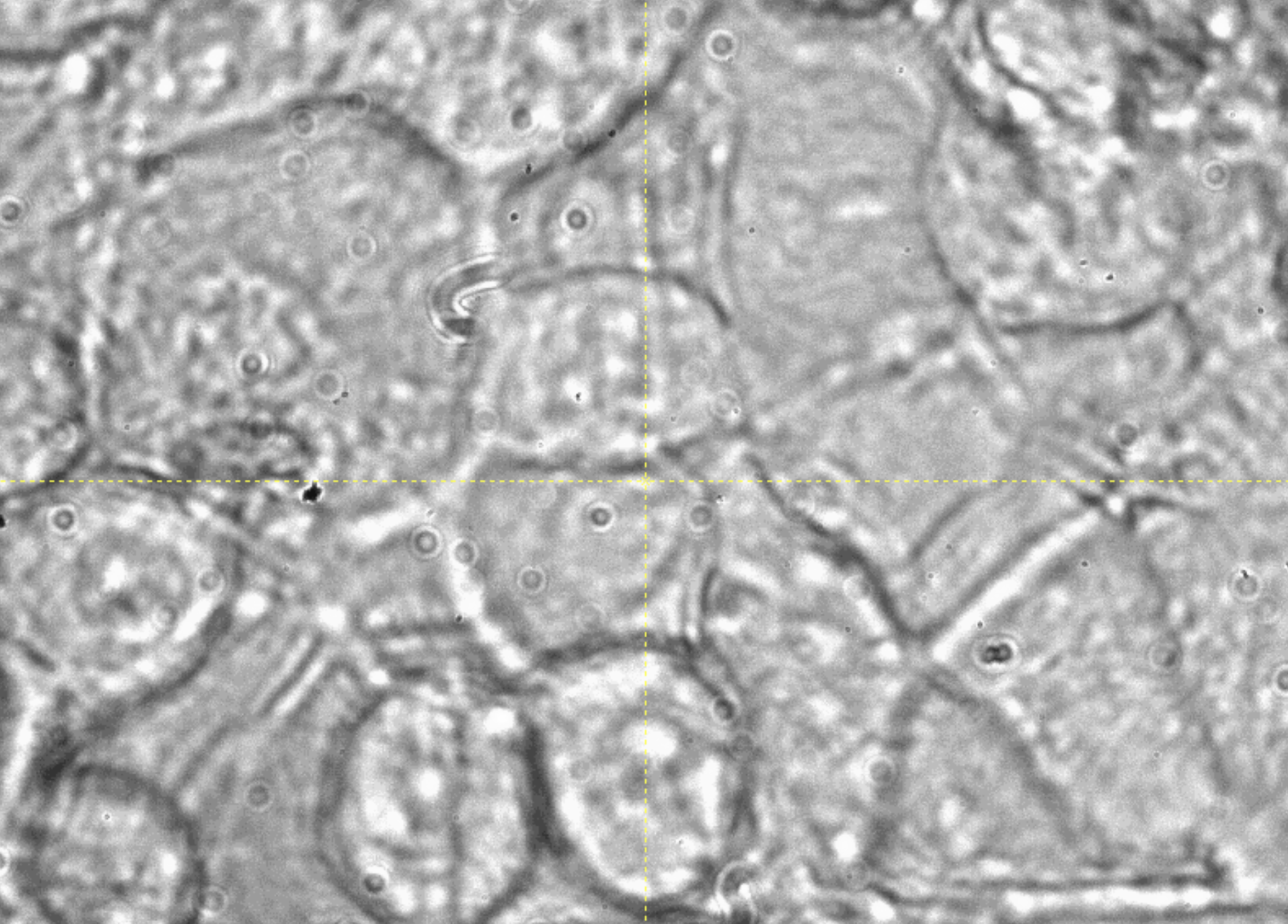}\\
  \includegraphics[width=0.2\textwidth]{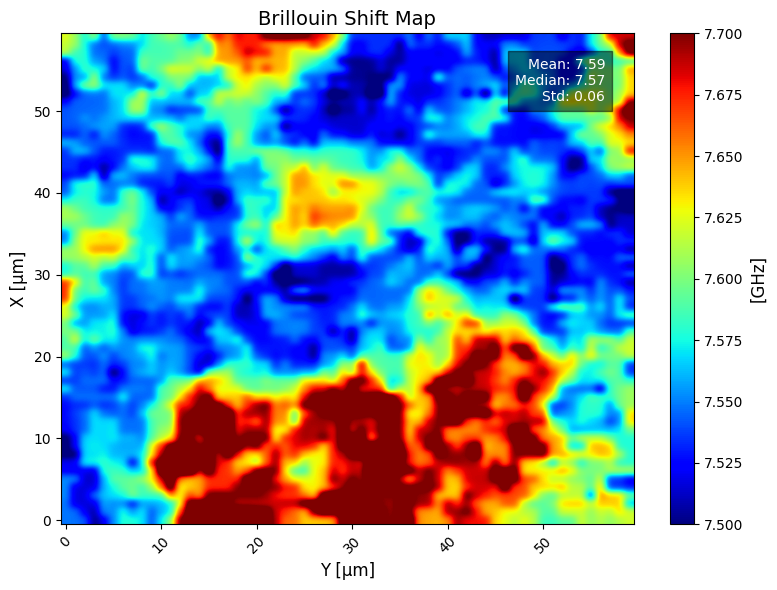}
  \includegraphics[width=0.2\textwidth]{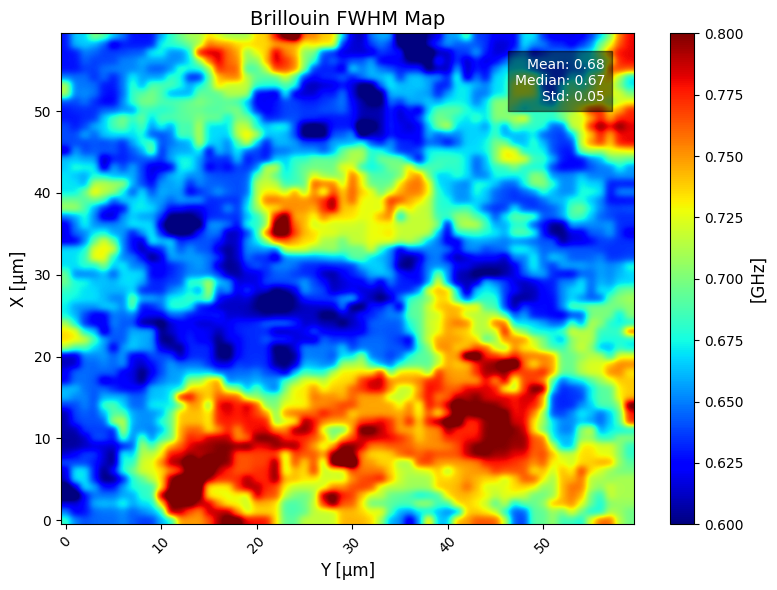}\\
  \includegraphics[width=0.2\textwidth]{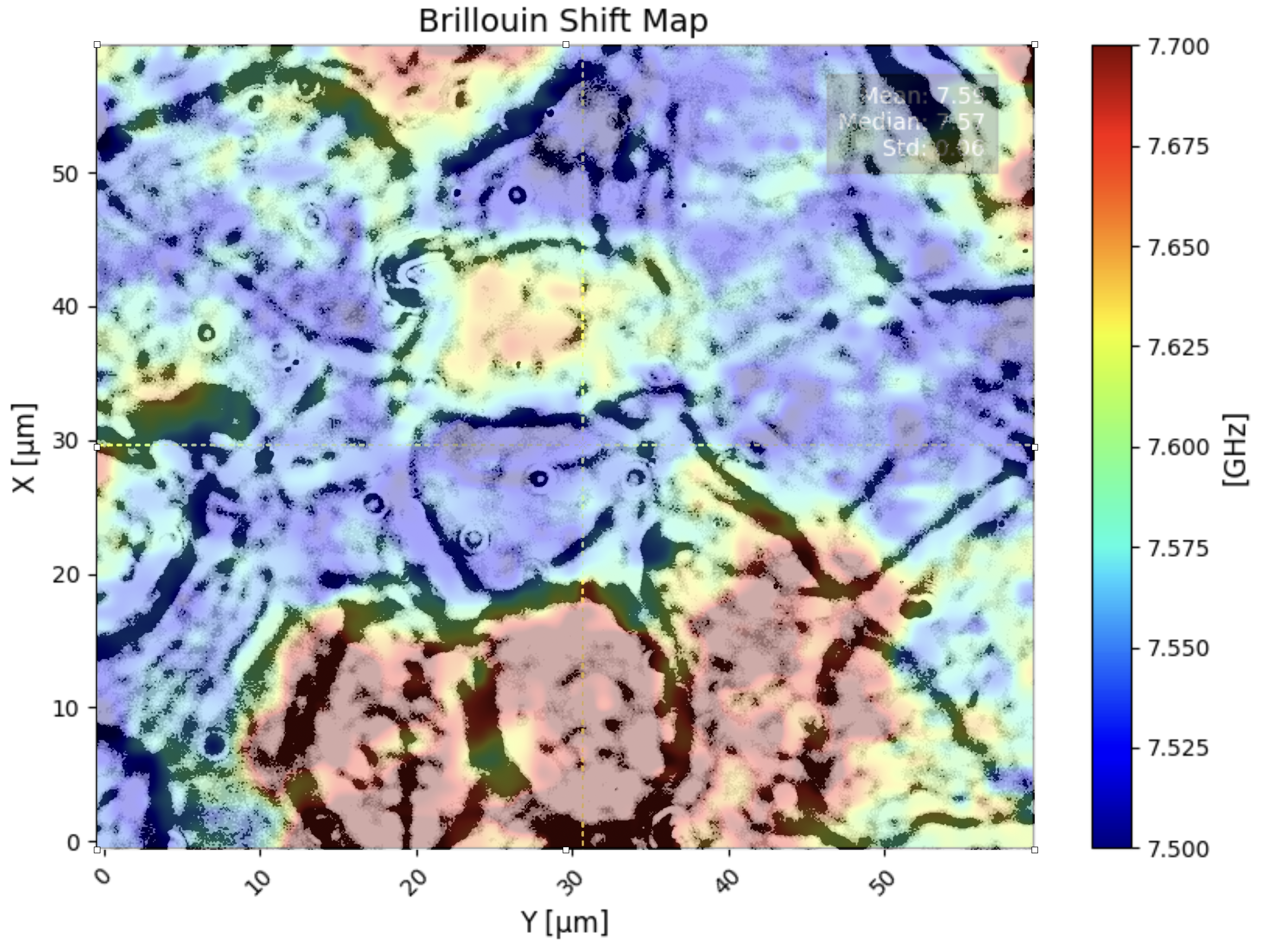}
  \includegraphics[width=0.2\textwidth]{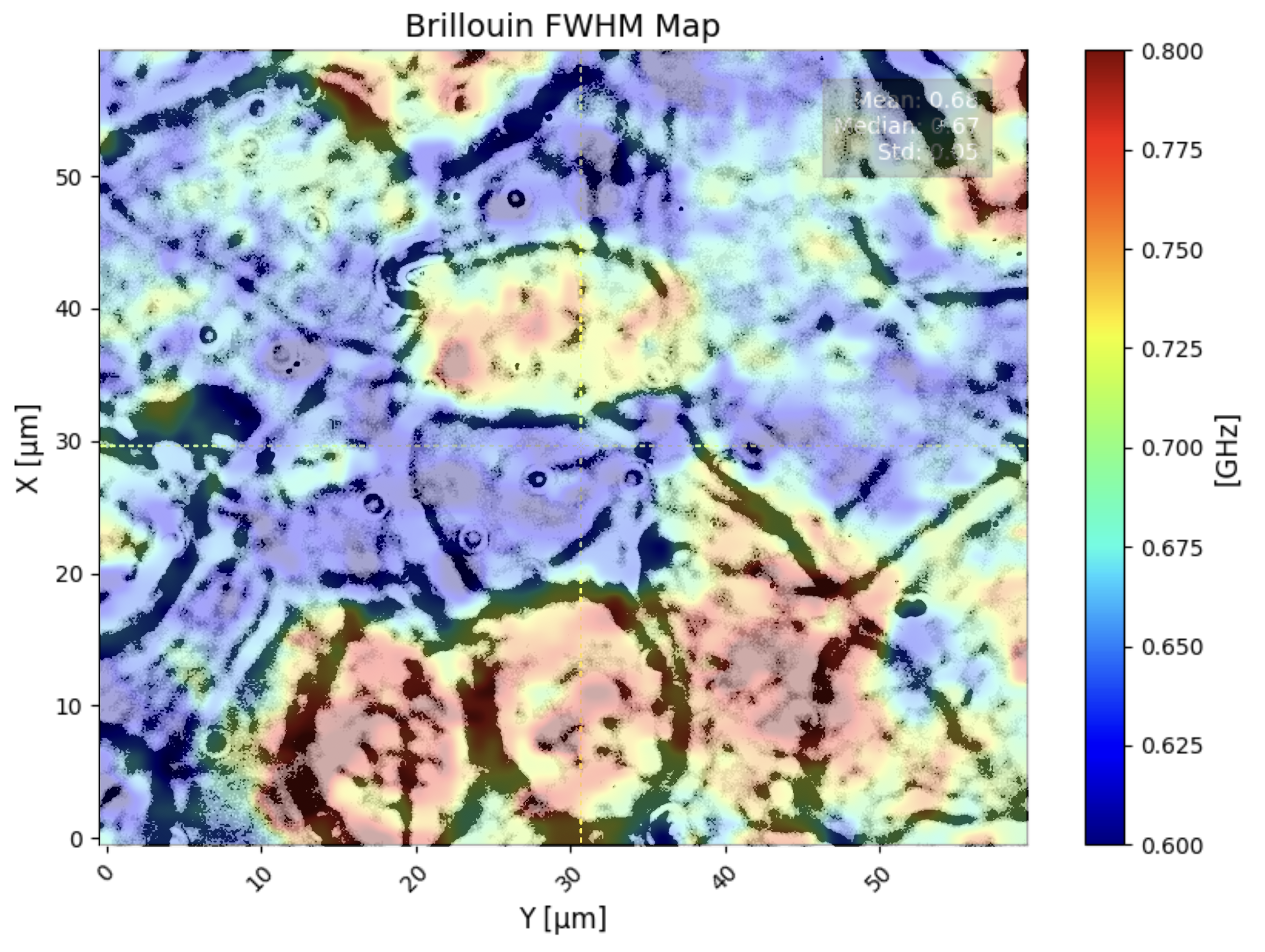}\\
  \includegraphics[width=0.2\textwidth]{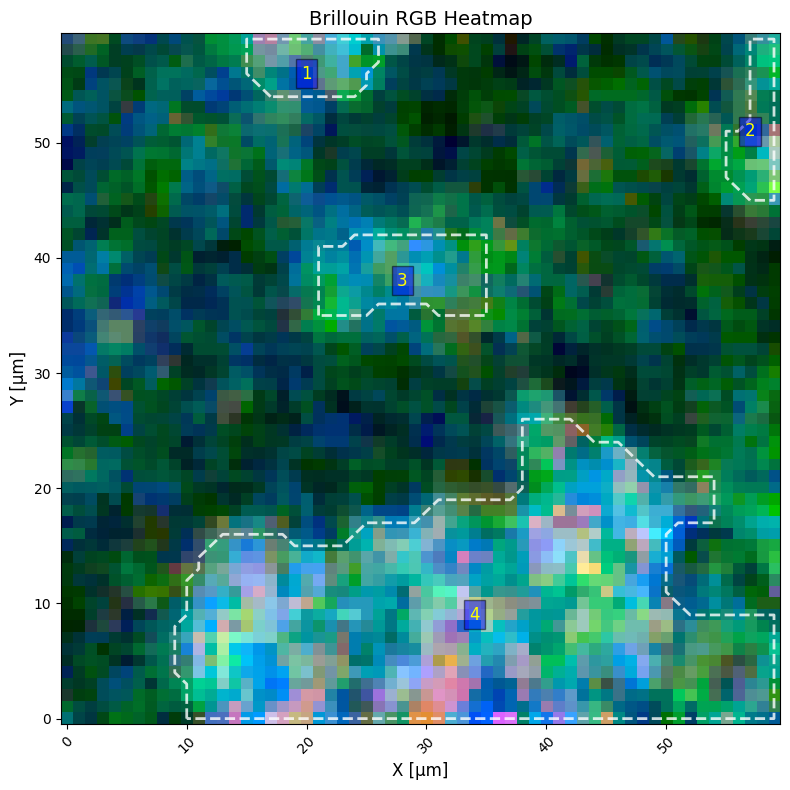}
  \caption{Top row - photo of the cell before and after laser exposure,  second row - Brillouin shift and FWHM heatmaps, third row - Brillouin shift and FWHM heatmaps blended with photo, last row - selected cells regions}
  \label{fig:mel_1}
\end{figure}

\begin{figure}[!h]
  \centering
  \includegraphics[width=0.4\textwidth]{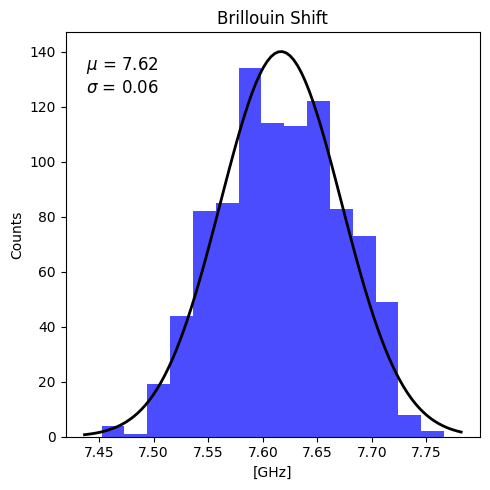}
\includegraphics[width=0.4\textwidth]{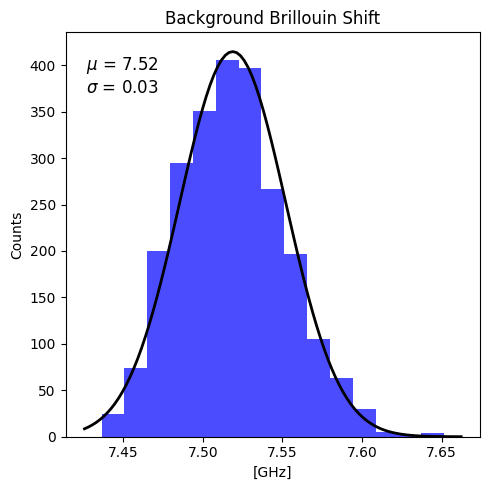}\\
\includegraphics[width=0.4\textwidth]{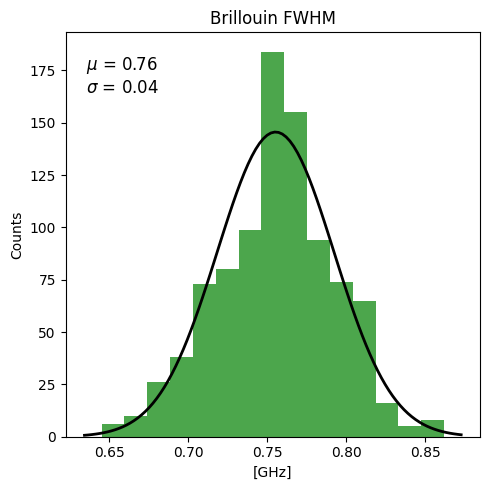}
\includegraphics[width=0.4\textwidth]{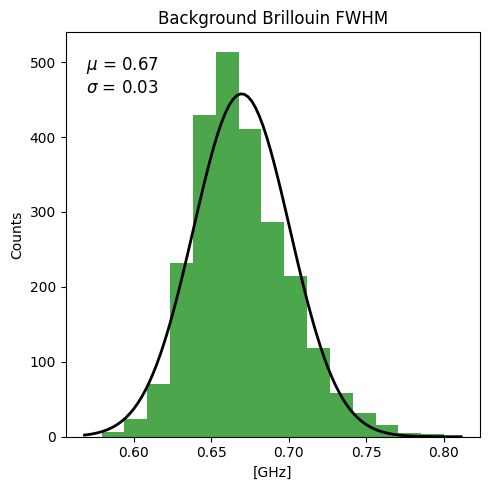}

  \caption{Top row - shift histogram of the cell group \#4 and background. Bottom row - FWHM histogram of the cell group \#4 and background.}
  \label{fig:mel_2}
\end{figure}

\begin{figure}[!h]
  \centering
  \includegraphics[width=0.2\textwidth]{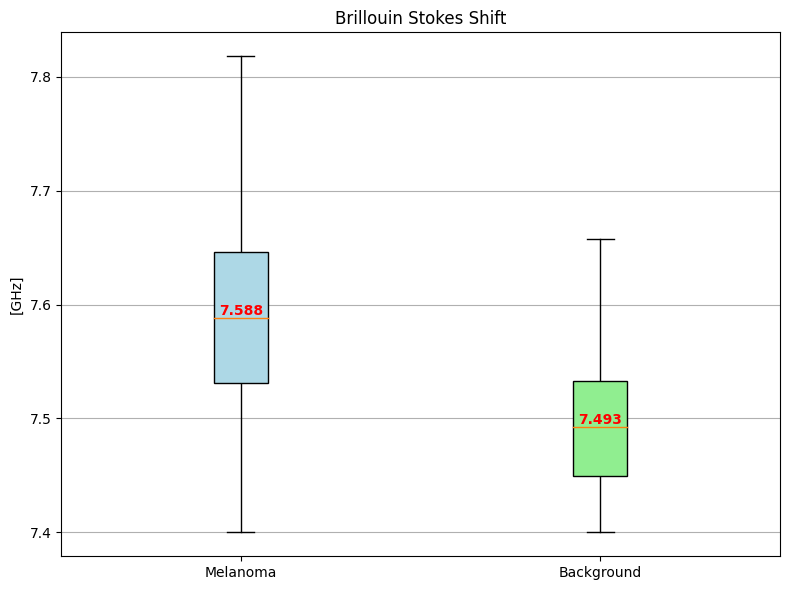}
  \includegraphics[width=0.2\textwidth]{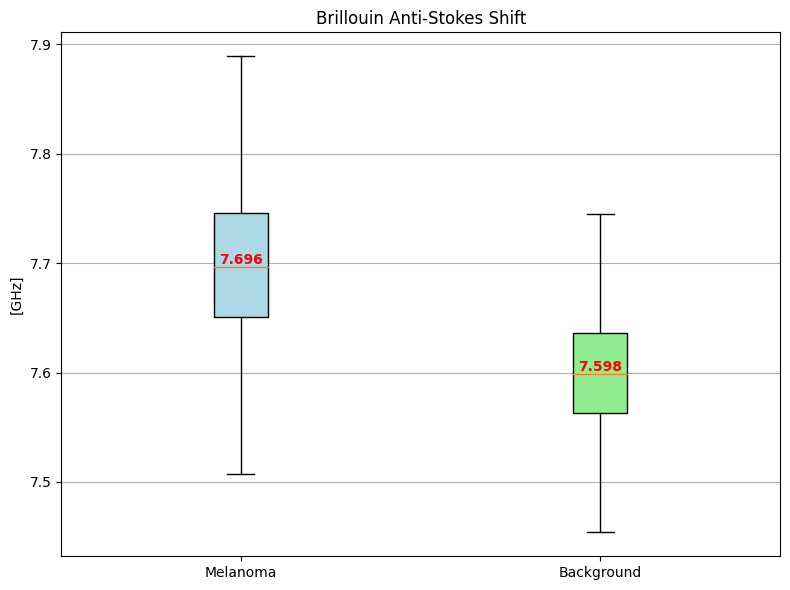}
  \includegraphics[width=0.2\textwidth]{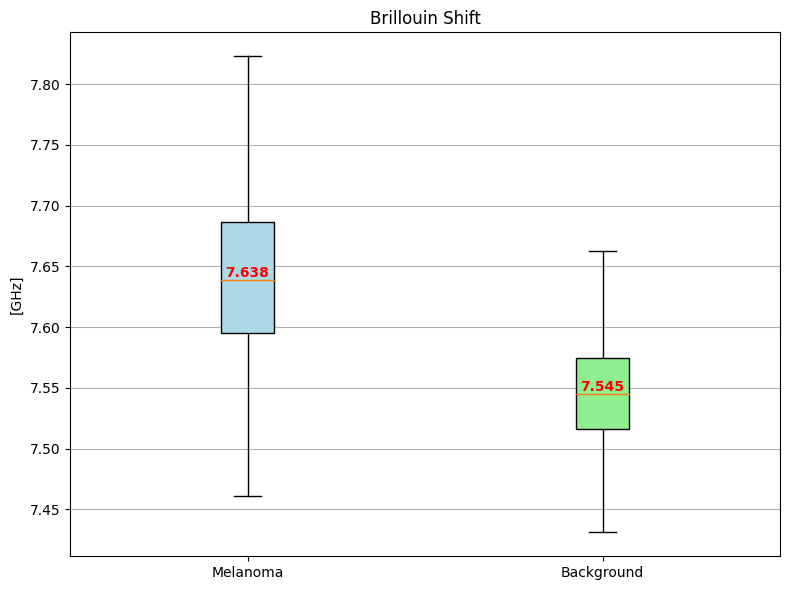}
  \\
  \includegraphics[width=0.2\textwidth]{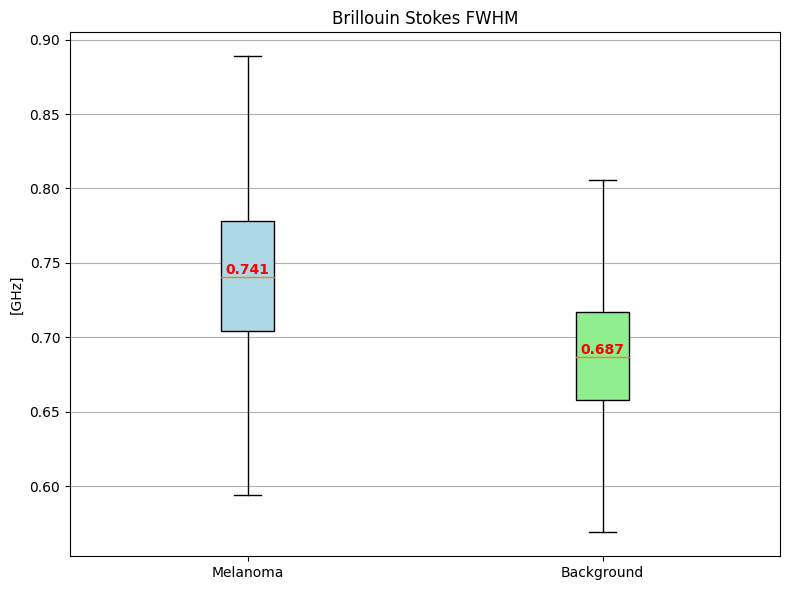}
  \includegraphics[width=0.2\textwidth]{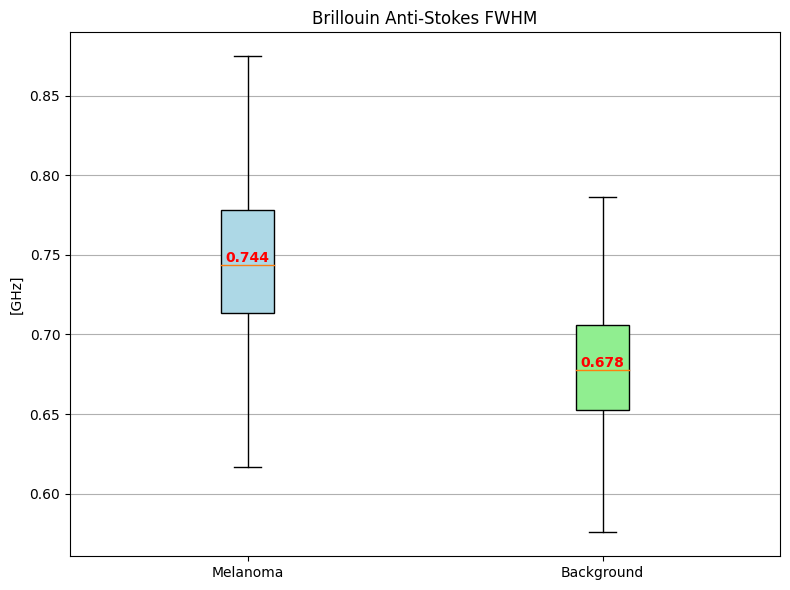}
  \includegraphics[width=0.2\textwidth]{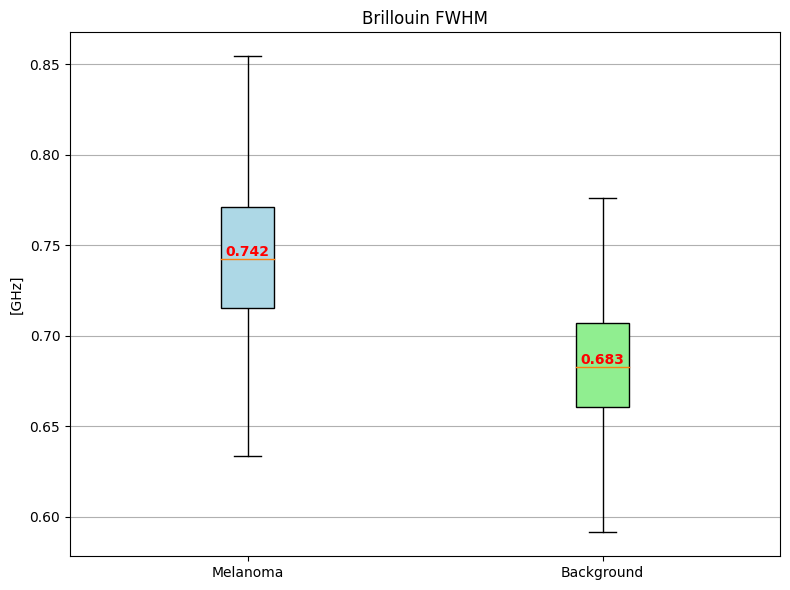}
  \caption{Top row - shift histograms of cells (stokes, anti-stokes and both), bottom row - FWHM histograms of cells (stokes, anti-stokes and both)\\
11 imagings (30 cells) in total.}
  \label{fig:mel_3}
\end{figure}

\begin{figure}[!h]
  \centering
  \includegraphics[width=0.2\textwidth]{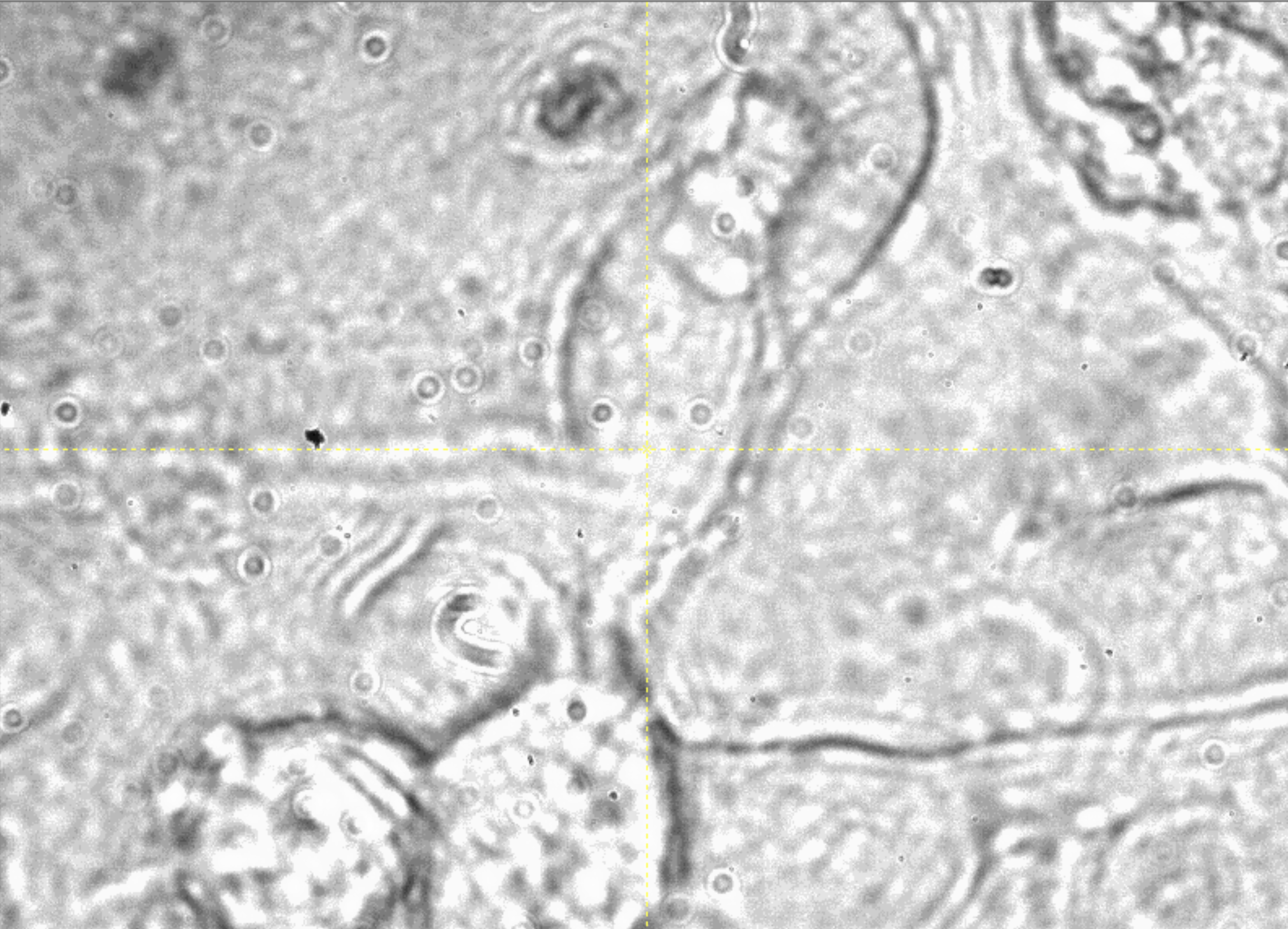}
  \includegraphics[width=0.2\textwidth]{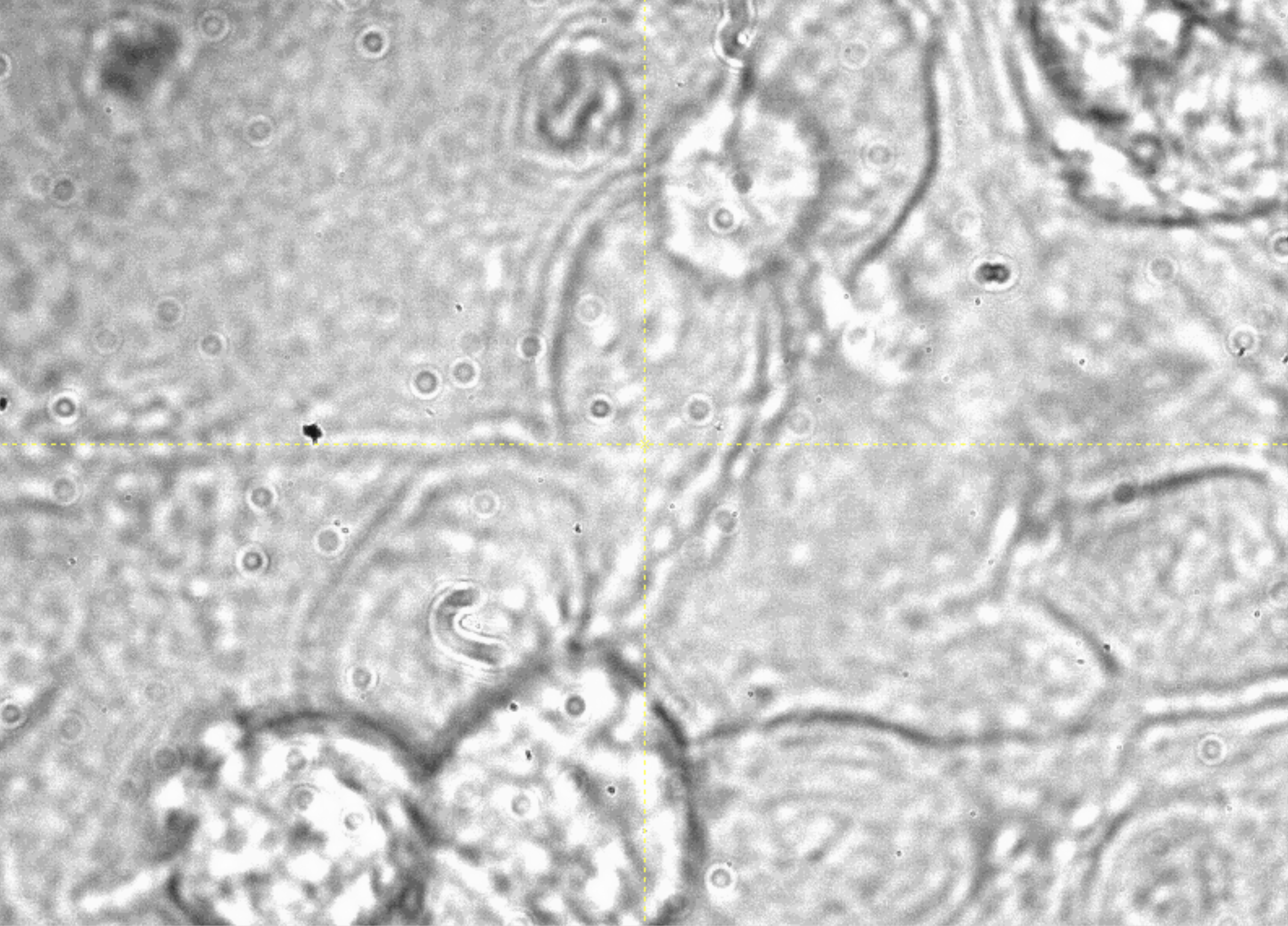}\\
  \includegraphics[width=0.2\textwidth]{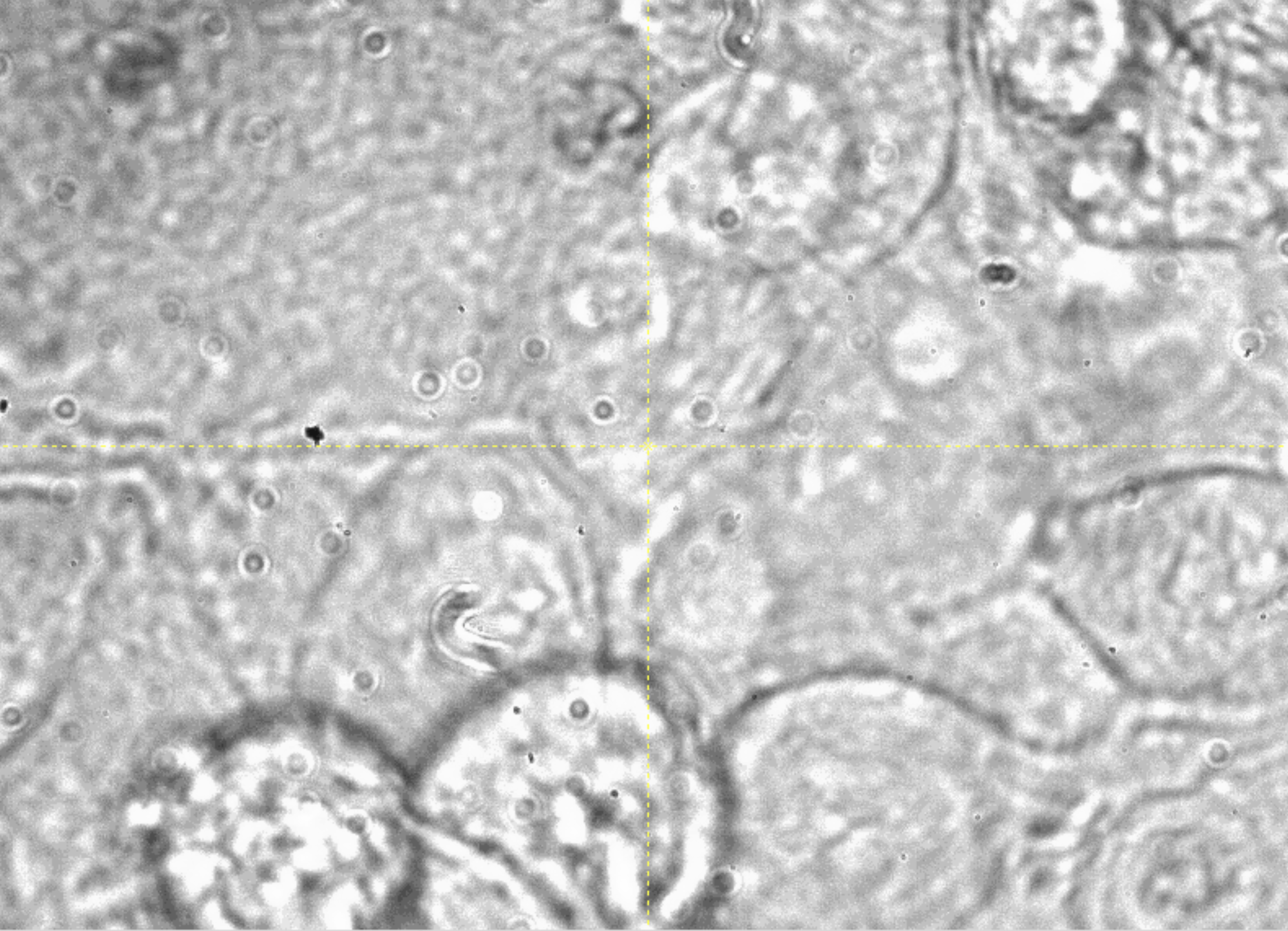}
  \includegraphics[width=0.2\textwidth]{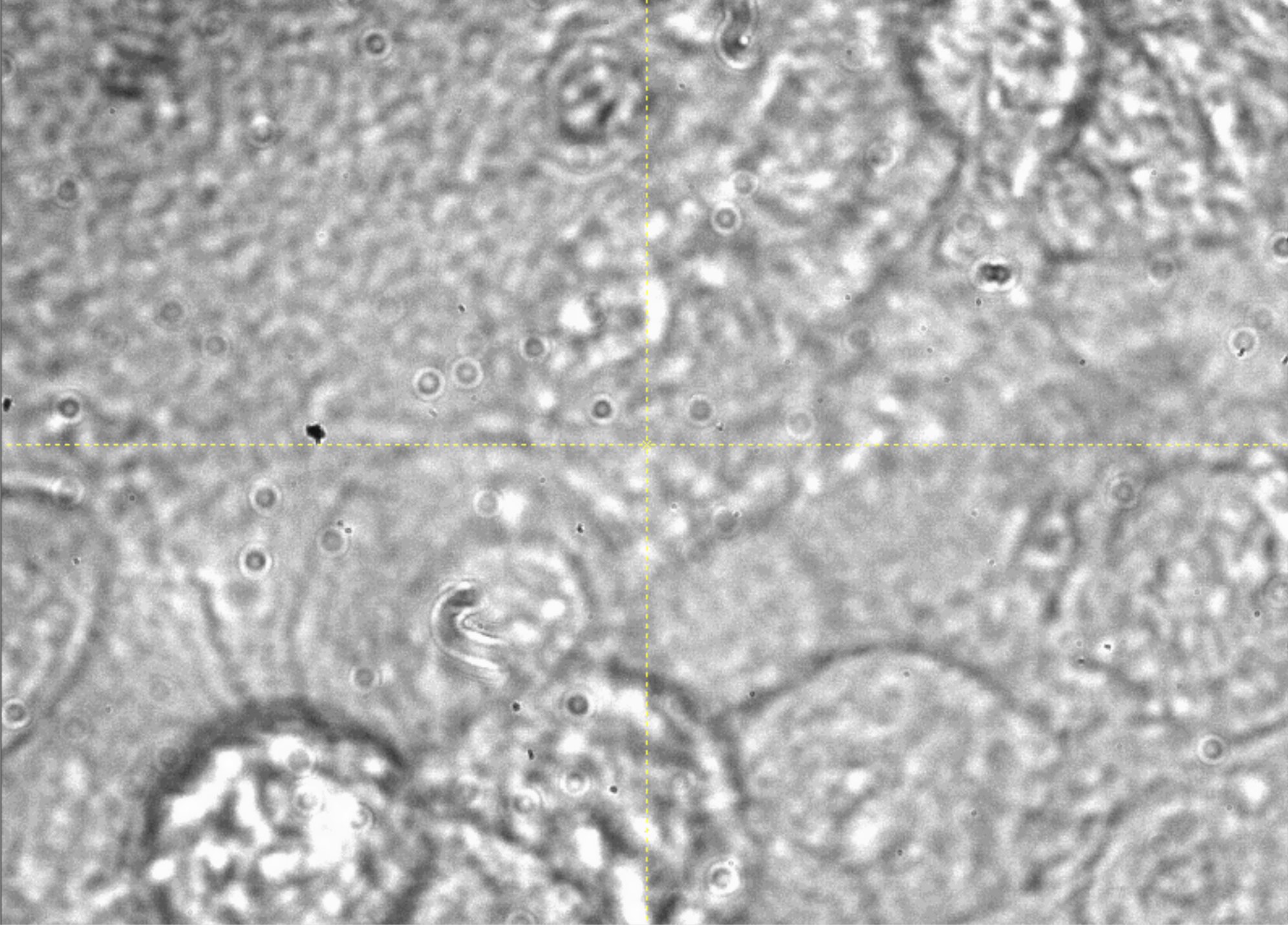}\\
  \includegraphics[width=0.2\textwidth]{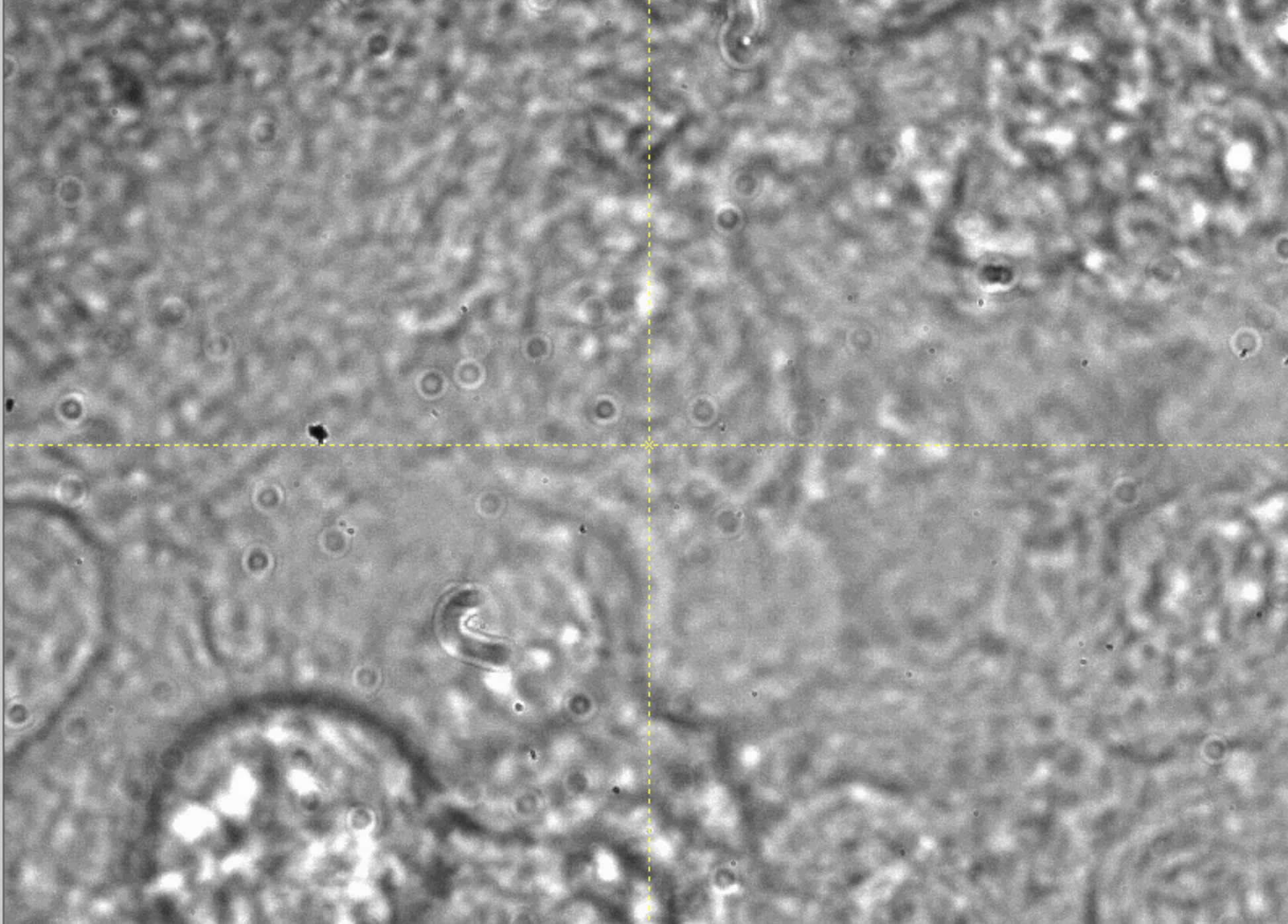}
  \includegraphics[width=0.2\textwidth]{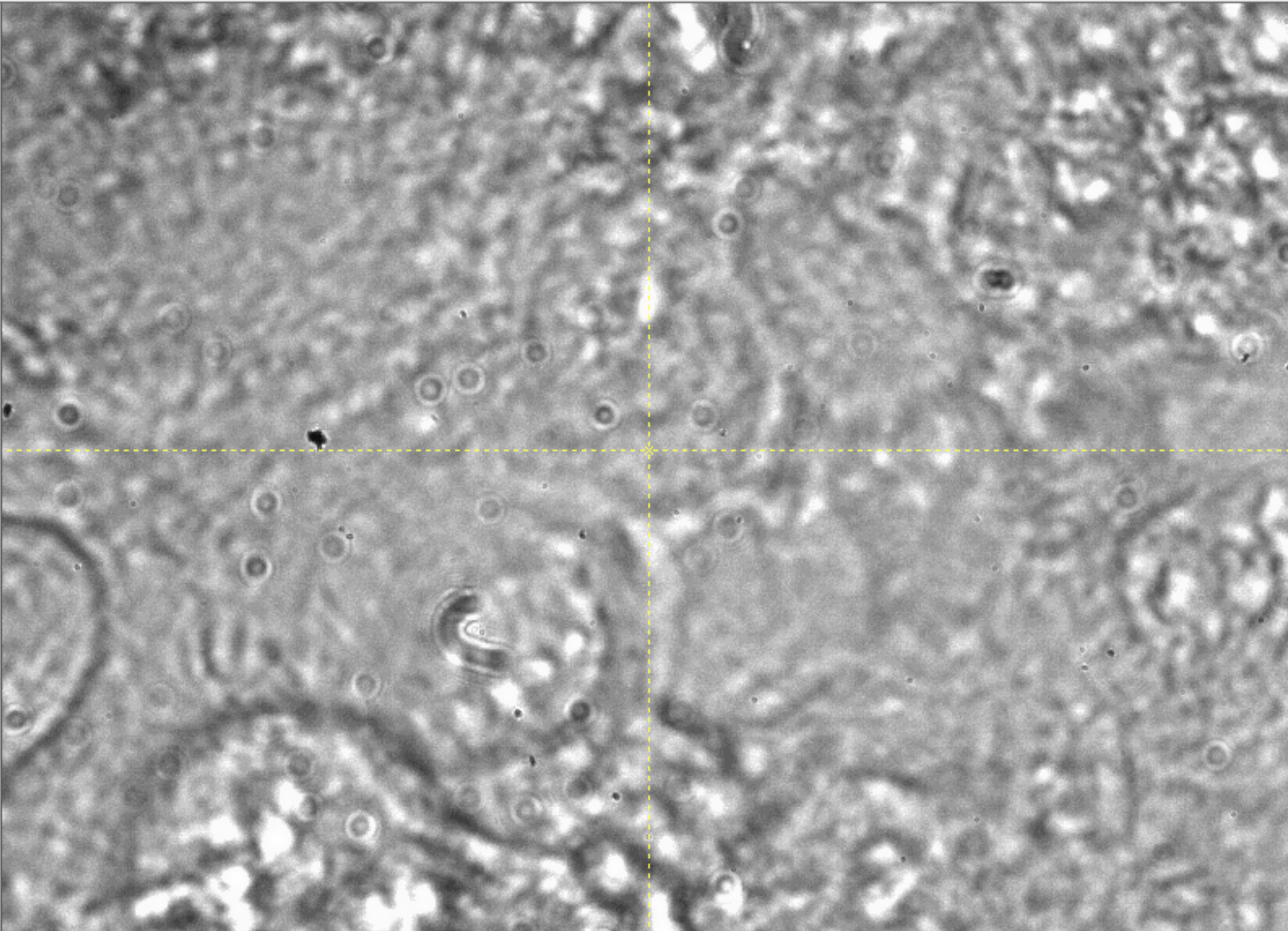}
  \caption{Top row - cell before and immediately after 6 minutes of laser exposure. Middle row - cell 10 and 20 minutes post-exposure, showing signs of degradation. Bottom row - cell 30 and 40 minutes post-exposure, where the cell undergoes complete destruction and annihilation.}
  \label{fig:cells}
\end{figure}

\section{Results and Discussion}

The Brillouin spectroscopy analysis revealed significant distinctions in both the Brillouin shift and the full width at half maximum (FWHM) between the melanoma cells and the surrounding background. Figures \ref{fig:mel_1}, \ref{fig:mel_2}, and \ref{fig:mel_3} showcase a series of heatmaps, histograms, and blended images, highlighting these differences.

\subsection{Brillouin Shift Observations}
The Brillouin shift values, mapped across the imaged samples, demonstrated that melanoma cells exhibited distinct mechanical properties when compared to their surrounding environment. This distinction is evident in the heatmaps (Figure \ref{fig:mel_1}), where higher Brillouin shift values were primarily associated with regions of cellular structures. Histograms (Figure \ref{fig:mel_2}) further support these findings, with cell group \#4 exhibiting a mean shift value that diverges from the background distribution. 

\subsection{Full Width at Half Maximum (FWHM) Analysis}
The FWHM values, reflective of the viscosity and structural heterogeneity within the cells, also presented measurable differences between the melanoma cells and the background. As demonstrated in Figure \ref{fig:mel_1}, the FWHM maps reveal that melanoma cells generally exhibit broader peaks, indicating greater internal complexity and mechanical inhomogeneity. The histograms for FWHM values (Figure \ref{fig:mel_2}) confirm this trend, with cell group \#4 showing a broader distribution compared to the background.

\subsection{Statistical Comparison and Implications}
The statistical analysis of the shift and FWHM values (Figure \ref{fig:mel_3}) confirmed a reproducible pattern across all 11 imaging runs. Notably, the anti-Stokes and Stokes components yielded consistent trends, demonstrating the robustness of the measurement. These findings suggest that Brillouin microscopy is a viable technique for distinguishing between melanoma cells and surrounding tissue based on their biomechanical properties.

\subsection{Impact of Laser Exposure}

The effect of laser exposure on melanoma cells was initially observed after an especially high-density (high-resolution) imaging run with the step size equal to 1$\mu$m with exposure (laser dwell time) equal to 40$\mu$s. After the exposure the cell has been observed for over 40 minutes, each subsequent observation occurring approximately every 10 minutes. shown  As shown in Figure \ref{fig:cells}, the cell undergoes visible morphological change following prolonged exposure.

Following the initial 6 minutes of laser exposure, significant morphological changes in the cells were observed, including early indicators of damage. Over the subsequent 10 to 20 minutes, the cells exhibited rapid rounding and a loss of amoeboid protrusions.

At the 30-minute mark, the cells reached a critical state where their outlines became indistinguishable, suggesting the onset of cell death. By 40 minutes, we inferred membrane rupture, as components of the cytoskeleton were observed to diffuse into the surrounding media. The absence of any discernible cell outline at this point indicated that the cells had been effectively destroyed due to prolonged laser exposure.

Notably, cells in close proximity to the imaged cell also exhibited altered morphology. This phenomenon may be attributed to both laser illumination and chemical signaling from the initially stressed cell \cite{czyz2018hgf}.

\section{Future direction}
The authors are aiming to investigate precise threshold for cell damage under high-power laser radiation. 
This observation highlights the potential cytotoxic effects of extended laser exposure, emphasizing the importance of optimizing laser parameters to avoid unwanted damage in live-cell imaging applications. The progressive degradation and eventual annihilation observed here could serve as a model for understanding photodamage mechanisms in melanoma cells and optimizing imaging protocols.

\appendix    

\acknowledgments 
 
This work was partially supported by the following grants:\\
AFOSR: FA9550-20-1-0366, FA9550-20-1-0367, FA9550-23-1-0599\\
NIH: 1R01GM127696, 1R21GM142107, 1R21CA269099\\
NASA/FDA: 80ARC023CA002E

\bibliography{report} 
\bibliographystyle{spiebib} 

\end{document}